# Optimized Deployment of Network Function for Resource Pooling Switch

Fei Hu, Jiong Du, Du Xu
Key Laboratory of Optical Fiber Sensing and Communications, Ministry of Education,
University of Electronic Science and Technology of China, Chengdu, P. R. China

*Abstract*—The disadvantages of the combination of traditional switches and middleboxes have being exposed under the condition of increasingly various network function demands, such as function flexibility, performance scalability and resource utilization. To solve this problem, we design Resource Pooling Switch Architecture (RPSA), which separates some non-essential functions from line card and allocate them in Network Function Pool (NFP) to provide flexible services for data plane in the form of Service Function Chains (SFC). As the performance of the whole system could be decided by whether function deployment is reasonable or not, we purpose heuristic algorithm called Modified Fiduccia-Mattheyses based Two Phase Algorithm (MFMTP) to optimize the deployment of functions. The simulation results show that this algorithm performs well in throughput and convergence.

*Keywords—Switch Architecture, Function Deployment, Network Function Virtualization, Service Function Chain*

## I. INTRODUCTION

In current network, Middleboxes, connecting to switches or routers, play an intrinsic and fundamental role in network security, overload detection, load balance, fault-tolerance and so on. Admittedly, this working mechanism provides high processing efficiency for network. But with the development of Internet, a wide range of new applications request more different and flexible function processing, the combination of traditional switches and middleboxes gradually show deficiency in meeting the requests of more complicated network. This is because for one thing, switches, encapsulated into closed "black box" usually perform badly in aspects of openness, expansiblity as well as inter-operability and the processing mode of pipeline in line card of switch in all possibility leads to waste of resources; for another, middleboxes, are usually expensive hardware with poor flexibility. This is why they are shifted from hardware to software with the help of Network Functions Virtualization (NFV). Taking both the disadvantages of traditional switches and middleboxes into account, we design RPSA based on NFV [1-4] and SFC to satisfy various network requests.

To mitigate the problem of resources utilization and take advantage of SFC to manage functions, we separate non-essential network functions from pipeline in line card and allocate them in NFP to provide flexible processing for packets. The general functions are kept as they used to be, conducting necessary procedures like integrity and logicality verification, After which, Classifier divides packets into different types according to matching rules to correspond to different SFCs in NFP. In term of this architecture, efficient function deployment and packets scheduling are two key issues to ensure performance of it. According to figure 1, Manager controls the whole process, including management of Network Functions (NFs), sending Service Function Path (SFP) manages to Classifier, and managing packet scheduling of Switch Fabric.

As for our knowledge, there exists no architecture like ours that not only could provide data plane of switch with flexible and customized network functions, but also could implement management of network functions and packet scheduling from level of architecture. Other existing researches only solve sub-problems [5-8].

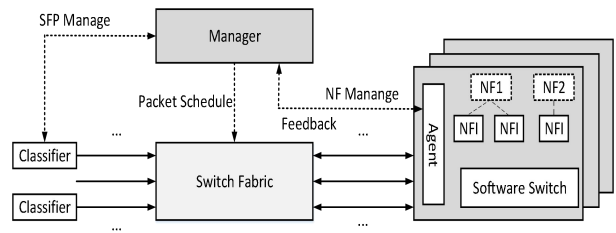

Fig. 1. Schematic Diagram of managing

In this paper, Firstly, we design RPSA architecture, which solves the problem of poor scalability and flexibility of switch by attaching NFV technology to it. RPSA could be of equal efficiency in function deployment and packet scheduling with powerful algorithms. As the packet scheduling problem has been addressed in our another literature [9], function deployment is the focus of this paper. We will use optimization theory to model this problem to Binary Integer Programming (BIP) and propose an heuristic algorithm called MFMTP. The simulation results prove its efficiency under this architecture.

The remainder of this paper is organized as follows. We introduce the key issue of function deployment under RPSA architecture and give an overview of the related work in section Ⅱ. Section Ⅲ is about the process of mathematical modeling. We discuss MFMTP algorithm and its time complexity in section Ⅳ. We present and analysis the simulation results in section Ⅴ. Last, we conclude this paper with final remark and perspectives for future work.

## II. RELATED WORK

The priorities of this architecture is obviously in two aspects. First of all, in the form of SFC and under the guidance of Manager, NFP could provide users with services whose composition, strategies and processing capabilities vary according to different requests. Furthermore, this architecture is able to improve utilization of resources by deep programmability and efficiently recycling and reusing existing resources.

To ensure those priorities, function deployment plays one of the most important roles in this architecture. We know from existing researches [5-12] that different network function deployment schemes have a significant impact on traffic processing performance as well as cost of system resources and under the constraint of resources, the optimal function deployment to meet all SFCs is very complicated and difficult.

When it comes to our architecture, efficient function deployment is especially important. As shown in figure 2, it is an example of one SFC in two deployment ways. In the first scheme, as dashed line A shows, after receiving all function processing in server 1, packets leave switching fabric. While as dashed line B shows, in scheme 2, after receiving $NFI_1$, $NFI_2$ and leaving server 2, packets need return to Switch Fabric and enter server 3 to receive $NFI_3$, after which, packets finally leave switching system.

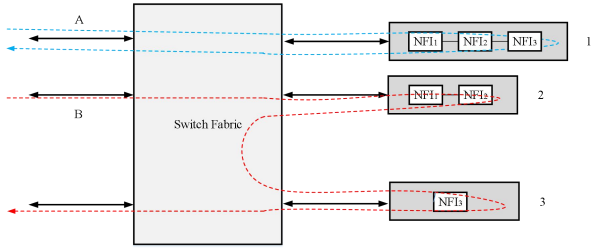

Fig. 2. Two Function Deployment Schemes

Comparing two schemes, we could conclude that the less the number of servers that traffic flows through, the less cost of physical links between Switch Fabric and NFP, the less delay and the higher throughput of switching system are. On the contrary, if one SFC requirement need more servers, traffic repeats round-trip transmission between Switch Fabric and NFP, which not only is a waste of limited resources, but also makes the direction of traffic disorder, the packets scheduling complicated and hence system performance poor. So, in our architecture, as the volume of inter-traffic between Switch Fabric and NFP reflects the performance of function deployment, to minimum inter-traffic is the optimization object in this paper.

The literature [7] proposes the VNF-CP (VNF chain placement) problem for maximizing the utilization of server node resources, while it is essentially different from our optimization goal. Literature [8] proposes heuristic algorithm, sets distance cost and establishment cost of NFV as optimization goal, to solve NFV-LP (NFV Location Problem), which is not fit to our architecture as we only consider distance cost. Literature [10] proposes Green algorithm to turn off as many free servers as possible, but it does not consider the distribution statement of traffic. Literature [11] focus on the problem of NFV deployment under mixed scenes, but it does not consider inter-traffic between servers. Literature [12] aims at minimum traffic cost, but it still need optimization. Since most algorithms, either need optimization themselves, or is not fit to our architecture, we will propose heuristic algorithm based on Fiduccia-Mattheyses (FM) algorithm [13], aiming at minimum inter-traffic between Switch Fabric and NFP, to ensure the performance of whole switching system.

## III. MATHEMATICAL MODELING

### A. System Model

Under RPSA architecture, packets correspond to specific SFCs, each of which represents bandwidth requirement of traffic, specific order of NFs and resource requirement of each NF. A specific SFC request can be expressed through SFC1 in Figure 3(a), during which the value above links between NF nodes represents bandwidth and the value in the rectangle on the node represents resource consumption of each NF. By merging the same NF in each SFC, such as the $NF_2$ contained in SFC1 and SFC2 in Figure 3(a), all of the SFC requests can be described as a weighted directed acyclic graph (SFC-Graph) $G^f = (V^f, E^f)$ shown in Figure 3(b). The vertex of the graph represents a network function; the resource consumption of corresponding NF is $C_f$; the edge $e_{nm} \in E^f$ of the graph indicates that there is traffic between the NFn and NFm and the bandwidth is $B(e_{nm})$.

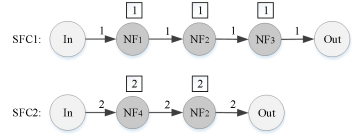 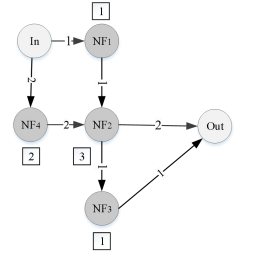

Fig. 3(a). SFC Request    Fig. 3(b). SFC-Graph

Fig. 3. SFC and SFC-Graphh

The standard of the IETF service function chain allows the same type of network functions to deploy multiple Network Function Instances (NFIs) to achieve high availability and load balancing. Based on SFC-Graph, the number of NFIs that each NF needs is calculated according to the load condition of each NF. Conditions that include the volume of total traffic arriving at NF, remaining resources of each server node, goals of load balancing and high availability. Then, based on the calculation result, SFC-Graph is transformed into weighted acyclic graph SFC-iGraph taking NFI as granularity. One example is shown in figure 4. Figure 4(a) represents the original SFC-Graph (we save edge weight for succinctness), and, after transforming and modulating toward links, SFC-iGraph is shown in figure 4(b).

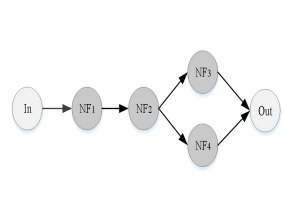 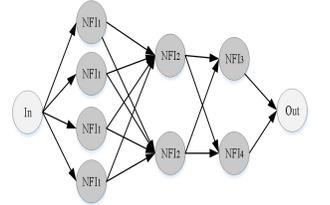

Fig. 4(a). SFC-Graph    Fig. 4(b). SFC-iGraph

Fig. 4  Illustration of SFC-Graph and SFC-iGraphh

In some cases, there are some relevance constraints on the traffic redistribution of input and output between adjacent NFs when split NFs. Those relevance constraints make it possible to optimize SFC-iGraph. One possible optimized result of figure 4(b) is as shown in figure 5. Comparing the two figures, we know that the linking condition between NFIs of adjacent NFs is more simple and traffic direction is more clear. So, in our architecture, the problem of function deployment under resources constraint equals to rationally deploying NFI nodes in the optimized SFC-iGraph to servers.

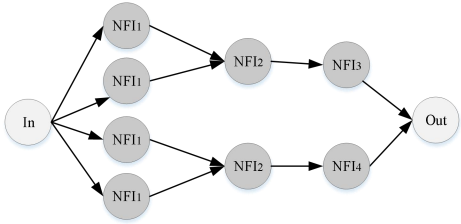

Fig. 5. Optimized SFC-iGraph

### B. Mathematical Modeling

In order to minimize the interaction traffic between Switch Farbric and NFP under the RPSA architecture, we adapt the optimization theory to model the above NF deployment problem into a 0-1 integer programming problem.

Firstly, the optimization goal is to minimize the inter-traffic between NFP and the Switch Fabric. In this function, $Cost_L$ represents the traffic cost between server and Switch Fabric. $U_{l_s^n}$ represents unit cost of link. And $Cost_{l_s^n}$ represents link resource consumption. So the objective function is:

$$\min Cost_L = \sum_{l_s^n \in L} Cost_{l_s^n} \cdot U_{l_s^n} \quad (1)$$

In this paper, we regard limited CPU capacity is the constraint of total resource of servers in NFP and assume that every server could offer same resource capacity. $N_f^i$ represents the number of NFI and $C_f^i$ represents resource consumption on every NFI. Therefore, the total consumption on all NFI should be:

$$\forall n \in N: \sum_{f \in F} y_n^{f_i} \cdot N_f^i \cdot C_f^i \leq T_n \quad (2)$$

Meanwhile, constraint of link bandwidth is shown as following. During which, $B_c$ represents requirement bandwidth of SFC.

$$\forall n \in N, \forall l_s^n \in L: \sum_{c \in n} \cdot B_c \leq B(l_s^n) \quad (3)$$

What's more, we use 0-1 integer variable $y_n^{f_i}$ to indicate whether one particular $f_i$ is deployed on n, and use 0-1 integer variable $x_n$ to indicate whether particular node n carries any NFI. $x_n$ and $y_n^{f_i}$ need to satisfy following variable constraints:

$$\forall f_i \in f: \sum_{n \in N} y_n^{f_i} = 1 \quad (4)$$

$$\forall n \in N: x_n \in \{0,1\} \quad (5)$$

$$\forall n \in N, \forall f_i \in F: y_n^{f_i} \in \{0,1\} \quad (6)$$

The relationship between $x_n$ and $y_n^{f_i}$ can be expressed as following:

$$\forall n \in N: x_n = \begin{cases} 0, & \sum_{f_i \in F} y_n^{f_i} = 0 \\ 1, & \sum_{f_i \in F} y_n^{f_i} > 0 \end{cases} \quad (7)$$

Finally, the number of service nodes is limited by the number of ports of switches:

$$\sum_{n \in N} x_n \leq P \quad (8)$$

## IV. MFMTP ALGORITHM

According to the 0-1 integer programming model proposed in section 3.2, we stipulate the network function deployment problem under the RPSA architecture as Graph Partition (GP) problem which is usually a NP-hard problem. Most existing researchers obtain solution of NP-hard problem through heuristic and approximation algorithm [13]. There is no polynomial time algorithm to solve the problem when $NP \neq P$. Therefore, based on Fiduccia-Mattheyses (FM) algorithm, we propose heuristic MFMTP algorithm to solve the problem of function deployment under RPSA architecture. FM algorithm and Relevancy Degree (RD), which will firstly be introduced in the remaining of this section are important to support MFMTP.

### A. Design of Algorithm

The FM algorithm is a widely used heuristic algorithm based on mobile iteration, which can effectively solve the problem of graph partitioning. Firstly, FM algorithm gives a set of initial divisions and the equilibrium condition in the dividing process and defines the reduction of objective function when one vertex is moved to another partition subset as vertex gain. Then it moves the point which has the maximum gain and satisfies equilibrium condition in every iteration to improve current graph partition. FM algorithm does not stop iterating until there is no more gain increase.

Because the typical application scenario of the FM algorithm [14], is not suitable for this structure and the vertex gain of NFIs in SFC-iGraph needs to reflect the inter-traffic between graph partitions (namely servers), this paper improves the FM algorithm to better adapt to the network function deployment under the RPSA architecture.

we define a new index RD to measure the flow relationship between NFIs in RPSA architecture.

Firstly, $C = (c_{i,j})$ is used to represent traffic matrix among NFI nodes, namely the edge weight of the directed acyclic SFC-iGraph. Under the condition of having already obtained initial function deployment solution, the relation between NFI node $n \in N$ and server $k$ is defined as external cost $E_{n,k}$, which could be described in following form:

$$\forall n \in N, n \notin k: E_{n,k} = \sum_{y \in k} c_{ny} \quad (9)$$

The internal cost $I_n$ of one NFI is defined as:

$$\forall n \in N, n \in k : I_n = \sum_{x \in k} c_{nx} \quad (10)$$

$RD_{n,k}$ is defined as the difference between $E_{n,k}$ and $I_n$:

$$RD_{n,k} = E_{n,k} - I_n \quad (11)$$

As shown in figure 6, taking $NFI_3$ and $NFI_4$ as an example could illustrate the conception of RD well. The RDs to indicate relevancy between $NFI_3$ and Server 1, between $NFI_3$ and Server 3 are *0* and *-30* respectively; the RDs to indicate relevancy between $NFI_4$ and Server 1, between $NFI_4$ and Server 2 are both *20*.

RD reflects the condition of inter-traffic between servers. The larger the $RD_{n,k}$ is, the larger vertex gain is if we move node n into server *k*. Therefore, the MFMTP algorithm uses RD as an important guideline for optimization and adjusting the initial deployment solution.

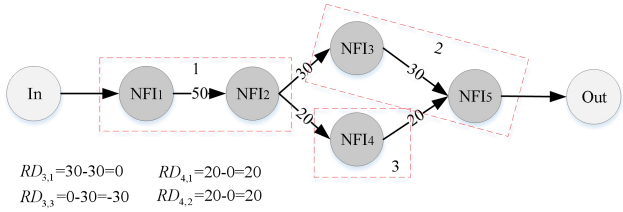

Fig. 6. Illustration of Relevancy Degree

MFMTP algorithm is designed as following two stages:
1) Initial deployment: Firstly MFMTP conducts Depth First Search (DFS) for SFC-iGraph. Then, considering the cost of servers, it selects server with minimum Remaining Resource Capacity (RRC) greedily for each NFI node which it has traversed under bandwidth constraint. If there are more than one server meeting the condition, First Come First Served (FCFS) deployment is adopted. After DFS traverses all the nodes of the SFC-iGraph, the algorithm gets a set of initial deployment solutions P, and each partition in the P represents a server
2) Optimization: Based on RD, MFMTP optimizes the initial deployment to realize minimum inter-traffic between Fabric Switch and NFP in this stage. If NFI satisfies the resource capacity of server and bandwidth, algorithm moves NFI nodes and destination servers in the order of RD from big to small and updates the RD of adjacent NFIs, which is a *move*. When the NFI is moved, it will be locked in this iteration until next iteration. Therefore, after the first iteration, all $RD_{n,k}$ of the NFI nodes are locked. MFMTP algorithm finds m *moves*, which correspond to maximum sum of $RD_{n,k}$ from all mobile nodes. It keeps the m *moves* of NFI and revokes the remaining $|V| - k$ *moves*, thus obtaining the optimal solution of P.

---

### *MFMTP Algorithm*

**Input**: $G_i = (V_i, E_i)$
**Output**: $Y = \{y_n^{f_i} | \forall f \in F, n \in N\}$, times
1: Initialization: *step=0, times = 1*
2: Do Depth First Search toward SFC-iGraph $G_i$, select server greedily to get initial solution
3: **while true do**
4:  Calculate $RD_{n,k}$ between NFI node *n* and initial solution $k(n \notin k)$, put $RD_{n,k}$ into max heap
5:  **while** $step < n$ **do**
6:   **if** $RRC_C(k) \geq C_n, RRC_{bw}(k) \geq BW_n$ **then**
7:    Update RD of adjacent NFIs
8:  **end while**
9:  Select $RD_{n,k}$ to maximum $G = \sum_{i=1}^{m} RD_{n,k}$
10:  **if** $G > 0$ **then**
11:   move NFI *n* to server *k, times ++*
12:  **else then break**
13: **end while**
14: **end**

### B. Analysis of Time Complexity

In this section, the time complexity of the MFMTP algorithm is analyzed.

First, the number of NFI nodes in SFC-iGraph is defined as *n*, the number of edges is *m* and the number of graph partitions (server numbers) obtained by the initial solution is *k*. At the initial deployment stage, time complexity of DFS for SFC-iGraph is $O(n*k+m)$.

At optimization stage, time complexity is decided by one iteration which is divided into three steps: (1) The first step is to get legal RD value from the maximum heap. It takes $O(n*k)$ to get data from heap. Under the worst condition, it is necessary to traverse all RDs, so the time complexity for one legal RD is $O(n*k*log_{10}(n*k))$. And considering that the maximum number of cycles is *n*, time complexity of the first step is $O(n^2*k*\lg(n*k))$; (2) The second step is to update all RDs of NFI nodes that are influenced by the update of $RD_{n,k}$. It is obviously that the upper limit of the number of NFI nodes equals to the degree of *n*. The time complexity for once update of general maximum heap is $O(log_{10}N)$, during which $N$ represents the number of elements in the heap, so the total time complexity for updating $RD_{n,k}$ is $O(\sum(D(n)*log_{10}(n*k)))$. $D(n)$ indicates the degree of node *n*. Considering that the sum of node degree is two times of edges, so the total time complexity for the second step of MFMTP algorithm is $O(m*log_{10}(n*k))$; 3) the third step is to choose the m biggest $RD_{n,k}$ and this step can be implemented by enumeration, so time complexity is $O(n)$.

To sum up, the total time complexity of the MFMTP algorithm is $O((n^2k+m)*log_{10}(n*k))$.

## V. SIMULATION AND ANALYSIS

### A. Environmental Setup

In this experiment, type of NF is set to 6 by investigating the commonly used NFs in the industry at present [15-16]. A SFC is randomly composed of [1,6] NFs and each NF is composed of [1,5] NFIs randomly. The traffic volume between NFIs in SFC-iGraph is distributed uniformly in [100,600], and the computing resource requirements for each NFI node are uniformly distributed in [100,600]. The computing resource and bandwidth resource of each server in NFP are 1000. In order to better show the performance of MFMTP algorithm, in the simulation, the FIRM algorithm is still used to address packet scheduling. Traffic arrives according to the Bernoulli process.

### B. Performance Index

In this paper, the following three indexes are used to verify the performance of MFMTP:
- NF interactive traffic: The inter-traffic between Switch Fabric and NFP. Less inter-traffic indicates less invalid transmission.
- Average iteration times：Average iteration times at the optimization stage is used to measure convergence rate of the algorithm.
- Throughput: larger throughout of the switching system indicates greater transmitting and data processing capacity of switching system. So the larger throughout is, the more reasonable the deployment of the NF is.

We compare MFMTP algorithm with Greedy for First Fit (GFF) algorithm [13]. GFF algorithm adopts the first fit strategy to meet the current resource and link constraint of each server node for NFI deployment.

### C. Simulation Results

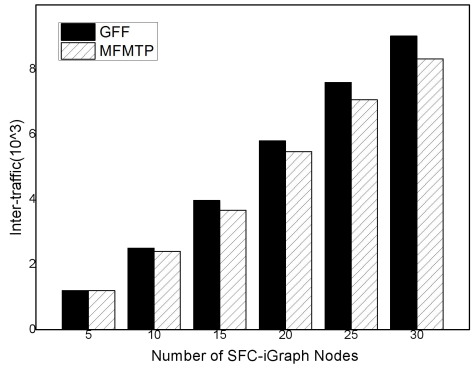

Fig. 7. NF Inter-traffic

The inter-traffic comparison results of MFMTP and GFF is shown in figure 7, from which we can firstly see that the average inter-traffic size of both algorithms become larger with the increase of NFI nodes in SFC-iGraph. In addition, we know that for SFC-iGraph with same number of NFI nodes, MFMTP performs better and the more nodes is , this priority is more obvious. When node number is larger than 20 (namely 20,25,30), compared with GFF, inter-traffic volume of MFMTP decreases 5.7%, 7.1% and 7.8% respectively. This is because under the condition of more complicated SFC-iGraph, MFMTP has more room to improve on the base of initial solutions

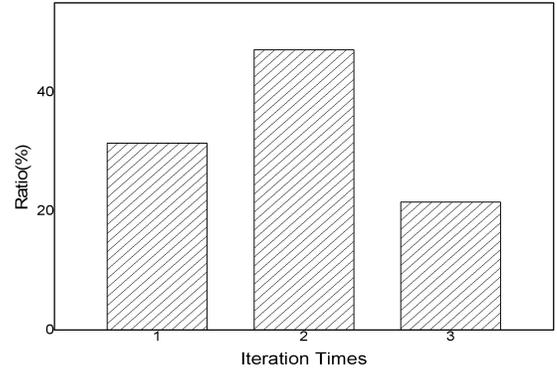

Fig. 8. Iteration Time of MFMTP under Different SFC-iGraph

Figure 8, the distribution of average iteration times of MFMTP algorithm under different SFC-iGraph structures, demonstrates that the MFMTP algorithm can complete the solution within three iterations (including the initial deployment where no optimization adjustment is needed, so the number of iteration is 1), which is in line with the theoretical expectation. Therefore, it is reasonable to conclude that MFMTP algorithm has good convergence for SFC-iGraph with different connections and traffic volume.

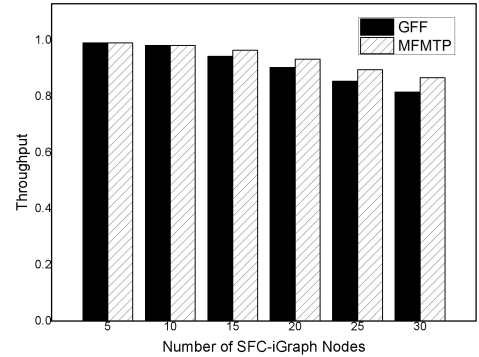

Fig. 9. Throughput under different SFC-iGraph Structure

Figure 9 compares the throughput index obtained by the MFMTP algorithm and GFF algorithm under different SFC-iGraph structures. It can be seen from the experimental results that the throughput performance of the MFMTP algorithm is better than that of GFF algorithm, even though under some specific conditions where the number nodes is small, they have near performance

Through the analysis of the simulation results, we know that, compared with the GFF algorithm, MFMTP algorithm makes the inter-traffic between Switch Fabric and NFP less, and performances better in throughput. Besides, MFMTP algorithm also has low time complexity and good convergence. Therefore, the MFMTP algorithm proposed in this chapter has strong application value.

## VI. CONCUSION

Considering that the increasingly different function requests exposes disadvantages of traditional switch and makes it almost impossible for the combination of traditional switching system and middleboxes to meet them. In this paper, first of all, based on NFV and SFC, we design RPSA, which could provide flexible and customized function processing for data plane by separating non-essential functions from pipeline in line card and allocating them in NFP. Furthermore, as function deployment is one of the most important issue in this architecture, we use optimization theory to model this problem to Binary Integer Programming (BIP) to optimize performance of this architecture and propose heuristic MFMTP algorithm. The simulation results show that this algorithm has good performance and considerable application value.

As perspectives for future work, we plan to expand this architecture to large-scale networking, such as 3-Clos where switch nodes could collaborate to provide resource reservation and path configure in advance for some traffic with high priority. We also plan to introduce the logically centralized control framework under the SDN architecture to facilitate the flexible and various deployments of the business choreography. Generally speaking, take advantage of SDN, applying our switch architecture to more complicated and practical network is mainly next task.

ACKNOWLEDGMENT

This work was supported in part by the National 863 Plan Program under Grant 2015AA016102.

REFERENCES

[1] Han B, Gopalakrishnan V, Ji L, Seungjoon Leee. Network function virtualization: Challenges and opportunities for innovation. IEEE Communications Magazine, 2015, 53(2): 90-97.
[2] R Jain, S Paul. "Network virtualization and software defined networking for cloud computing: a survey," IEEE Communications Magazine, 2013, 51(11): 24-31.
[3] J Matias, J Garay, N Toledo, J Unzilla, E Jacob. "Toward an SDN-enabled NFV architecture," IEEE Communications Magazine, 2015, 53(4): 187-193.
[4] S V Rossem, W Tavernier, D Coll, M Pickavet, P Demeesteret. "Introducing Development Features for Virtualized Network Services," IEEE Communications Magazine, 2018, PP(99):2-10.
[5] D Kliazovich, P Bouvry, S U Khan. "DENS: data center energy-efficient network-aware scheduling," Cluster computing. Hangzhou, China 2013, 16(1): 65-75.
[6] S Clayman, E Maini, A Galis, A Manzalini, N Mazzocca. The dynamic placement of virtual network functions. Network Operations and Management Symposium (NOMS). Krakow, Poland, 2014: 1-9.
[7] Q Zhang, T Xiao, F Liu. "A Manzalini, N Mazzoccal. Joint Optimization of Chain Placement and Request Scheduling for Network Function Virtualization," Distributed Computing Systems (ICDCS), 2017 37th International Conference on. IEEE. Krakow, Poland 2017: 731-741.
[8] R Cohen, L Lewin-Eytan, J S Naor, D Raz. "Near optimal placement of virtual network functions," Computer Communications (INFOCOM), 2015 IEEE Conference . Kowloon, Hong Kong, 2015: 1346-1354.
[9] http://confy.eai.eu/#manage-paper/id/276134/cid/52486/tid/1909
[10] M Xia, M Shirazipour, Y Zhang, H Green, A Takacs. Network function deployment for NFV chaining in packet/optical datacenters. Journal of Lightwave Technology, 2015, 33(8): 1565-1570.
[11] H Moens, F D Turc. "VNF-P: A model for efficient placement of virtualized network functions," International Conference on Network and Service Management. Rio de Janeiro, Brazil, IEEE, 2014:418-423
[12] P W Chi, Y C Huang. "Lei C L. Efficient NFV deployment in data center networks," Communications (ICC), 2015 IEEE International Conference on. IEEE. London, UK, 2015: 5290-5295.
[13] K Andreev, H Racke. "Balanced graph partitionin," Theory of Computing Systems, 2006, 39(6): 929-939.
[14] C M Fiduccia, R M Mattheyses. "A linear-time heuristic for improving network partitions," Papers on Twenty-five years of electronic design automation. ACM, 1988: 241-247.
[15] Y Li, M Chen. "Software-defined network function virtualization: A survey," IEEE Access, 2015, 3: 2542-2553.
[16] W John, K Pentikousis, G Agapiou, E Jacob, M Kind. "A Manzalini. Research directions in network service chaining," Future Networks and Services (SDN4FNS), IEEE SDN. Trento, Italy, 2013: 1-7.